# Non-perturbative Renormalization of Lattice Operators

A. VLADIKAS

Dip. di Fisica, Univ. di Roma "Tor Vergata"

INFN, Sezione di Roma II,

Via della Ricerca Scientifica 1,

I-00133 Rome,

Italy.

September 1995

We briefly review and compare three methods (one perturbative, one based on Ward Identities and one non-perturbative) for the calculation of the renormalization constants of lattice operators. The following results are presented: (a) non perturbative renormalization of the operators with light quarks; (b) the renormalization constants with a heavy (charm) quark mass and its KLM improvement; (c) the non perturbative determination of the mixing of the $\Delta S = 2$ operator.

# Non-perturbative Renormalization of Lattice Operators

A. Vladikas [a]

[a]Dipartimento di Fisica, Università di Roma 'Tor Vergata' and INFN, Sezione di Roma II,
Via della Ricerca Scientifica 1, I-00133 Rome, Italy.

We briefly review and compare three methods (one perturbative, one based on Ward Identities and one non-perturbative) for the calculation of the renormalization constants of lattice operators. The following results are presented: (a) non perturbative renormalization of the operators with light quarks; (b) the renormalization constants with a heavy (charm) quark mass and its KLM improvement; (c) the non perturbative determination of the mixing of the $\Delta S = 2$ operator.

## 1. Introduction

The Renormalization Constants (RC's) of lattice operators are a necessary ingredient in the calculation of physical Matrix Elements (ME) from lattice correlation functions. Schematically, a physical amplitude $A_{\alpha \to \beta}$ of a transition $\alpha \to \beta$ is given by

$$A_{\alpha \to \beta} = C_W(\mu) Z_O(a\mu) \langle \alpha | O(a) | \beta \rangle \qquad (1)$$

$C_W$ is the Wilson coefficient of the OPE, calculated in Perturbation Theory (PT) at a renormalization scale $\mu$; $\langle \alpha | O | \beta \rangle$ is the lattice bare ME of the operator relevant to the physical process, simulated numerically; a is the lattice spacing; $Z_O$ is the operator's renormalization constant. It can be calculated either in PT, or, non perturbatively, through Ward Identities (WI) or with a recently proposed non-perturbative (NP) method (Sects. 2 - 4). This review is focused on the systematic errors which afflict these calculations (Sects. 5 - 8) It is limited to Wilson and Clover 2- and 4-fermion operators.

## 2. Evaluation of RC's from PT

The perturbative renormalization of lattice operators will be reviewed elsewhere [1]; here we only highlight the essential features of the method. Consider a generic bilinear bare fermionic operator $O(a) = \bar\psi \Gamma \psi$ and a generic 1PI Green function $\Gamma_O(p, g_0, a) = \langle p | O_\Gamma | p \rangle$ ($g_0$ is the bare coupling and $p$ stands for the momenta of the external states). We will denote by $\hat O(\mu)$ and $\hat\Gamma_O(p, g^2, \mu)$ the finite renormalized counterparts of the above quantities and by $\Gamma_O^{(0)}$ the tree level estimate of $\Gamma_O$. The RC $Z_O(a\mu)$ is then defined through $\hat\Gamma_O(p, g^2, \mu) = Z_O(g_0^2, a\mu) \Gamma_O(p, g_0^2, a)$. It depends on the choice of regularization and renormalization scheme. The 1-loop estimate of $\Gamma_O$ is

$$\Gamma_O = [1 + \frac{g_0^2}{(4\pi)^2}(\gamma \ln(pa)^2 + C_{LAT})]\Gamma_O^{(0)} \qquad (2)$$

where $\gamma$ is the anomalous dimension and $C_{LAT}$ a finite coefficient. The renormalization condition usually imposed on $\hat\Gamma_O$ in lattice perturbative calculations (e.g. [2]) is the one chosen in the corresponding perturbative calculation in the continuum; normally this is the $\overline{MS}$ renormalization scheme in a Dimensional Regularization (DR), such as Dimensional Reduction, 't Hooft - Veltman etc. Thus we impose

$$lim_{a \to 0} \hat\Gamma_O |_{p^2 = \mu^2} = [1 + \frac{g^2}{(4\pi)^2} C_{DR}]\Gamma_O^{(0)} \qquad (3)$$

where $C_{DR}$ is the finite coefficient of the 1-loop calculation in the continuum (i.e. analogous to $C_{LAT}$ in eq.(2)). This is a choice of convenience: the Wilson coefficients $C_W$ of eq.(1) are known in $\overline{MS}$ (e.g. [3]). The RC's must consequently be in $\overline{MS}$ if the l.h.s. of eq.(1) is to be Renormalization Group invariant. The 1-loop perturbative estimate of the RC then follows:

$$Z_O = 1 - \frac{g_0^2}{(4\pi)^2}[\gamma \ln(\mu a)^2 + C_{LAT} - C_{DR}] \qquad (4)$$

The dependence of $Z_O$ on $C_{DR}$ comes from the choice of the $\overline{MS}$ renormalization scheme,

whereas its dependence on $C_{LAT}$ from the choice of lattice regularization. Two perturbative calculations are thus necessary, one in the continuum for $C_{DR}$ and one on the lattice for $C_{LAT}$.

The following points must be stressed: (1) The reliability of the perturbative calculation rests on the assumption that we work with a large enough cutoff; i.e. $\Lambda_{QCD} \ll a^{-1}$. (2) The renormalization condition of eq.(3) is defined in the limit $a \to 0$; i.e. all $\mathcal{O}(a)$ terms (and higher) vanish. (3) Operators which are protected by WI's, such as the vector current $V_\mu = \bar\psi \gamma_\mu \psi$, have $\gamma = 0$ and $C_{DR} = 0$ (for the axial current $A_\mu = \bar\psi \gamma_\mu \gamma_5 \psi$, a finite renormalization is neccessary in order to impose the WI's). In this case the $\overline{MS}$ renormalization condition of eq.(3) coincides with the one used in the NP method (see Sect. 4).

The calculation of the RC's in 1-loop PT suffers from $\mathcal{O}(g_0^4)$ systematic errors. Moreover, the lattice coupling $g_0^2$ is known to be a bad expansion parameter [4]. One particularly simple remedy consists in the choice of the boosted coupling $\tilde g_0^2 = g_0^2/U_\Box$ ($U_\Box$ stands for the average plaquette) as a better expansion parameter. This we call Boosted Perturbation Theory (BPT). Note that in theory, when working at 1-loop, it is immaterial which coupling (e.g. $g_0^2, g_{\overline{MS}}^2, \tilde g_0^2$) is used in eq.(4). This ambiguity in the choice of coupling is lifted in (and up to) 2-loop PT [5].

Another subtlety is that Standard lattice PT (SPT) ignores finite volume effects, since loop momenta are integrated rather than summed. Recently [6] this was taken into account by summing over discrete momenta; this is known as Discrete SPT (DSPT). Combining BPT with Discrete PT gives the RC's in Discrete Boosted PT (DBPT).

The PT method is reliable for finite and logarithmically divergent operators, the renormalization of which does not require a subtraction of lower dimensional operators. In the case of mixing with operators with lower dimension, a non-perturbative subtraction is necessary. This is because any non perturbative contribution to the mixing coefficients, being $\sim \exp[-\int^{g_0} d\bar g / \beta(\bar g)]$, when multiplied by the divergent behaviour $a^{-1}$, will give rise to an $a^{-1} \exp[-\int^{g_0} d\bar g / \beta(\bar g)] \sim \Lambda_{QCD}$ term, which will not vanish in the continuum limit $a \to 0$. This is the most serious limitation of the perturbative method.

### 3. Evaluation of RC's from chiral WI's

The most accurate method for a non perturbative determination of RC's is based on chiral WI's. Only a limited number of RC's can be extracted from them, such as $Z_V$, $Z_A$ and the ratio $Z_P/Z_S$ of the RC's of the pseudoscalar ($P = \bar\psi \gamma_5 \psi$) and scalar ($S = \bar\psi \psi$) densities. The RC's are determined by requiring that bare correlation functions calculated (numerically) on the lattice, when renormalized, obey the WI's of the theory. For example, the WI related to the vector flavour symmetry on the lattice gives rise to a point split conserved current $V_\mu^C$ [7], for which $Z_{V^C} = 1$. Since the local vector current is not conserved on the lattice, we fix its renormalization by requiring that the renormalized current $\hat V_\mu = Z_V V_\mu$ obey the same vector WI. This implies that $Z_V$ can be determined from a ratio of ME's of the kind

$$Z_V = \frac{\langle \alpha | V_\mu^C | \beta \rangle}{\langle \alpha | V_\mu | \beta \rangle} \qquad (5)$$

For the axial current and the ratio $P/S$ a similar procedure can be applied, provided that the breaking of axial symmetry by Wilson fermions is properly restored at vanishing lattice cutoff [7]. If the calculation is performed with WI's on hadronic states, everything is explicitly gauge invariant. If the external states are quark states, gauge fixing is necessary and the presence of Gribov copies may in principle undermine the method.

### 4. Non-perturbative evaluation of RC's

Recently, the following new non-perturbative (NP) method, based on the renormalization of quark state correlation functions has been proposed [6]: the quark state correlation function $G_O(x,y) = \langle 0 | \psi(x) O_\Gamma(0) \bar\psi(y) | 0 \rangle$ is calculated numerically in coordinate space at fixed coupling $\beta$ and hopping parameter K. Its forward amputated Fourier Transform, $\Lambda_O(p)$, is then calculated. From it, by a suitable projection in Dirac indices, $\check{P}$, we obtain the projected correlation function $\Gamma_O = Tr[\check{P}\Lambda_O]$, on which the following (off-shell) renormalization condition is imposed

$$Z_O(a\mu, g_0^2) Z_\psi^{-1}(a\mu, g_0^2) \Gamma_O(pa)|_{p^2=\mu^2} = 1 \qquad (6)$$

| Method   | $Z_V$(W) - Ref. | $Z_V$(SW) - Ref. |
|----------|-----------------|------------------|
| SPT      | 0.83 [2]        | 0.90 [11]        |
| BPT      | 0.71 [4]        | 0.83 [4]         |
| WI(2-pt) | 0.57(2) [8]     | 0.82(1) [9]      |
| WI(3-pt) | 0.72(3) [8]     | 0.82(1) [9]      |

Table 1
$Z_V$ calculated in SPT, BPT and with the WI method from ratios of $\langle V_\mu V_\mu \rangle$ (2-pt) and $\langle P V_\mu P \rangle$ (3-pt) correlation functions with the Clover(SW) and Wilson(W) actions.

If we know $Z_\psi$ we can solve eq.(6) for $Z_O$. The best determination of $Z_\psi$ is from the same eq.(6) with $O = V_\mu^C$, for which $Z_{V^C} = 1$ (equivalently we can use $O = V_\mu$ and $Z_V$ from of eq.(5)). Another estimate of $Z_\psi$ can be obtained from a lattice version of the propagator's derivative "$\partial S^{-1}/\partial p$" (see [6]). Note that the result depends on the momentum of the external states and on the gauge. The Landau gauge is a good choice.

In the case of $V_\mu$ and $A_\mu$, the renormalization condition of eq.(6) defines renormalized currents which automatically obey the Current Algebra (see [6] for the proof and some practical restrictions). It should also be noted that $C_{DR} = 0$ for the ME's of the currents and thus the two renormalization conditions eq.(3) and eq.(6) coincide. For all other operators, the renormalization conditions being different, we still need to match the RC computed from eq.(6) to the continuum RC of eq.(3). This matching has to be computed in continuum PT. Thus with the NP method we avoid lattice PT with its problem of a bad expansion parameter.

## 5. Discretization Errors

At finite lattice spacing, operators mix with higher dimensional (irrelevant) operators. This mixing spoils the multiplicativity of the renormalization of the lattice operators, which is only recovered in the limit $a \to 0$. The contribution of higher dimensional operators to the numerical calculation of RC's (from WI's or NP-ly) depends on the ME's used. This dependence signals the presence of systematic errors which are effectively $\mathcal{O}(ap)$ in the scaling limit. With the Clover action, the RC's of improved operators (in the spirit of [10]) have effectively $\mathcal{O}(g_0^2 ap)$ errors. In order to control these systematic errors we must ensure that $\mu \ll \mathcal{O}(a^{-1})$. Since we also need a good control of non perturbative effects, working in a window

$$\Lambda_{QCD} \ll \mu \ll \mathcal{O}(a^{-1}) \qquad (7)$$

is an essential ingredient for the reliability of any renormalization procedure performed at finite cutoff (e.g. BPT).

In conclusion, when the condition of eq.(7) is satisfied, the estimates of the RC's from 1-loop PT (good up to $\mathcal{O}(g_0^4)$) and those obtained from different ME's either with the NP or with the WI method (good to $\mathcal{O}(ap)$ in the Wilson case and $\mathcal{O}(g_0^2 ap)$ in the Clover one) should all agree. In present day calculations at $\beta = 6.0$, however, such discrepancies are 20% - 30% for Wilson fermions. They reduce to 5% - 10% for light Clover fermions (see Table 1). Notice that BPT also improves the agreement. This good control of the systematic errors with the Clover action at current $\beta$ values can worsen, however, as the quark mass increases. This case will be considered in detail in Sect. 7

## 6. Light Quarks

Results from WI estimates for the Wilson and Clover action for quark bilinear operators have been around for some time (see Table 1). Last year the first NP results have been presented with the Clover action ($\beta = 6.0$, $K = 0.1425$, $V = 16^3 \times 32$ and 36 confs.) [6],[12]. The estimates of $Z_V$ and $Z_S$ obtained from BSPT, BDPT, NP and WI (the latter only for $Z_V$) were in good agreement in a window of $\mu^2 a^2$ values. The situation was less satisfactory for $Z_A$, where lattice artifacts were seen. For $Z_P$ a discrepancy between the NP estimate and those from BSPT and BDPT signals large 2-loop corrections.

In this conference the first results for the same operators at the same $V$ and $\beta$ values ($K = 0.1515, 0.1530, 0.1550,$ and 125 confs.) were presented for the Wilson case [13]. A good agreement between the SPT and NP estimates is seen in a large window in $\mu^2 a^2$ (Figs 1,2 at $K = 0.1530$). The $Z_A$ results (not shown here) are of similar good quality, whereas $Z_S$ (also not shown) fluctuates more (just like in the Clover data of [6]). These Wilson results (but obtained on a subset of 36 confs.), compared to the Clover ones of [6], are significantly less fluctuating. This may be attributed to the fact that the Clover action and the improved operators are less local than the Wilson ones.

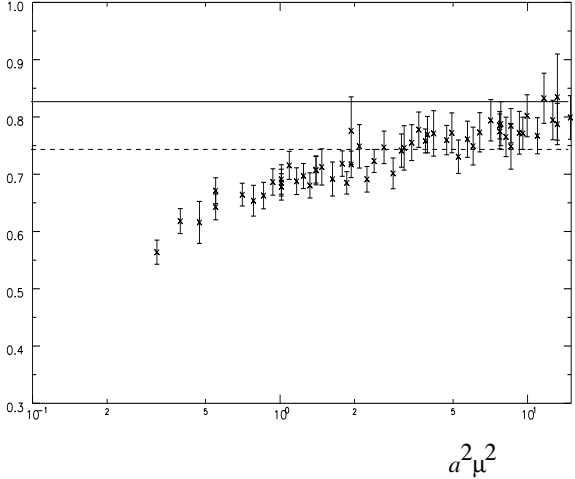

Figure 1. Wilson action estimate of $Z_V$ from NP method (points), from PT (continuous line) and from tadpole improved PT (dashed line).

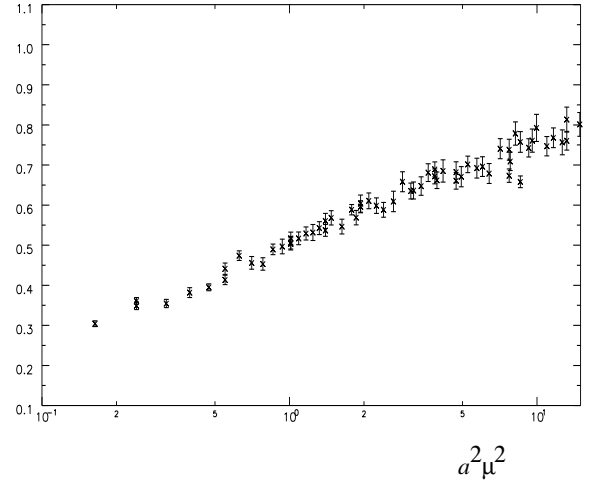

Figure 2. Wilson action NP estimate of $Z_P$

The same collaboration has also calculated nucleon structure functions [13], [14]. This requires the renormalization of operators like $O_{\{\mu\nu\}}$ and $O_{\{\mu\nu\sigma\}}$, from which the 1st and 2nd moments of the structure functions are obtained. For example, $O_{\{\mu\nu\}}$ is defined by

$$O_{\{\mu\nu\}} = \bar{\psi}\gamma_{\{\mu}D_{\nu\}}\psi - traces \qquad (8)$$

where $\{...\}$ stands for symmetrization of the Lorentz indices. The PT estimates of these RC's with the Wilson and Clover actions have been obtained in [15] and extended to higher rank operators in [14] for the Wilson case.

The RC of $O_{\{\mu\nu\}}$ ($K = 0.1530$) is given in Fig. 3; similar results have been obtained for the other relevant operators. Again a satisfactory window in $\mu^2 a^2$ is visible, in which the operator's anomalous dimension is clearly seen in the slope of the curve.

So far we have assumed that the Gribov ambiguity does not introduce any errors. This assumption may be questioned in the light of recent results which demonstrated that: (1) at least some Gribov copies are lattice artifacts [16],[17] and (2) they spoil the measurement of some gauge dependent correlations [16]. In [18] the influence of Gribov copies was explicitly studied by evaluating $Z_A$, on the same ensemble of 36 confs. ($\beta = 6.0$, $V = 16^3 \times 32$, $K = 0.1425$), both from a gauge independent WI on hadron states and a gauge dependent one on quark states. In the latter case $Z_A$ was computed on several Gribov copies. The result

$$Z_A = 1.06(2) \ gauge \ ind.$$
$$Z_A = 1.08(5) \ gauge \ dep. \qquad (9)$$

supports our assumption. We will assume that this result generalizes to all RC's calculated either

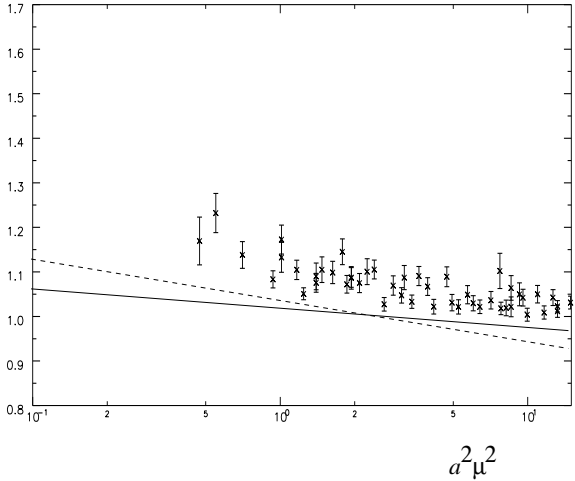

Figure 3. Same as Fig. 1 for the RC of the operator $O_{\{\mu\nu\}}$

from quark state WI's or from the NP method.

## 7. Heavy quarks

As the quark mass increases towards the charm (e.g. in semileptonic decays) the systematic error of $\mathcal{O}(am)$ in Wilson fermions ($\mathcal{O}(g_0^2 am)$ in Clover fermions) is bound to grow beyond control. In [4], [19] a remedy has been proposed in the form of the so-called KLM normalization factors. These arise when a tree level lattice object (propagator, correlation function etc) is matched to its continuum counterpart at zero spatial momentum $\vec{p} = \vec{0}$. For example consider the continuum tree level 2-quark correlation function of the local vector current in 3-momentum space ($\vec{p} = \vec{0}$):

$$C_{CON}^{V_\mu} = \int d^3x \int d^3y \langle 0 | \psi_1(x) V_\mu(0) \bar{\psi}_2(y) | 0 \rangle \quad (10)$$

where 1 and 2 label distinct flavours. In terms of its lattice counterpart $C_{LAT}^{V_\mu}(\vec{p} = \vec{0}, t)$, calculated with Wilson fermions, it is given by

$$C_{CON}^{V_\mu} = 2K_1 \, 2K_2 (1 + am_1)(1 + am_2) C_{LAT}^{V_\mu} \quad (11)$$

Besides the traditional $2K$ quark field normalizations, there are the extra $(1 + am)$ KLM normalizations. They differ from 1 at $a \neq 0$. The above relationship is also valid when we match $C_{CON}^{V_k}$ with $C_{LAT}^{V_k}$, the quark correlation function of the spatial components of the conserved current $V_k^C$ ($k = 1, 2, 3$). For the temporal case $V_0^C$ there is an extra factor, as observed by Bernard [20]:

$$C_{CON}^{V_0^C} = 2K_1 \, 2K_2 (1 + am_1)(1 + am_2)$$
$$\frac{2}{2 + am_1 + am_2} C_{LAT}^{V_0^C} \quad (12)$$

The claim of [4], [19] is that taking these KLM corrections into account when calculating, say, $Z_V$ from eq.(5), will soften its "mass dependence" (which is really a systematic $\mathcal{O}(am)$ error). A counter-claim raised in [21] is that since KLM factors only remove the $\mathcal{O}(am)$ error at tree level, they might at best cause a softer "mass dependence" of the RC's obtained at a given, fixed ME. They will not, however, eliminate the $\mathcal{O}(am)$ systematic error which shows up as dependence of the $Z_O$ estimate on the ME's it was obtained from. Also, since the KLM factors are calculated at $\vec{p} = \vec{0}$, they do not correct $\mathcal{O}(a|\vec{p}|)$ effects.

In the results shown below [22], the value of the quark mass used in the KLM factors is simply taken to be $am = 1/(2K) - 1/(2K_c)$ with $K_c$ extracted non perturbatively. Using, instead, the original Mean Field Tadpole Improved (MFTI) quark mass of [4] gives similar results [21], [23].

In Fig 4 we reproduce from [21] ($\beta = 6.4, V = 24^3 \times 60$, 20 confs.) the "mass dependence" of the ratio

$$R_1 = \frac{\langle K | V_0^C | D \rangle}{\langle K | V_0^L | D \rangle} \quad (13)$$

(obtained from 3-point correlation functions) without and with KLM corrections. It is clear that the KLM claim is well supported. Next we consider the ratios of 2- and 3-point correlations

$$R_2 = \frac{\langle 0 | V_k^C | \rho \rangle}{\langle 0 | V_k^L | \rho \rangle} \quad \text{and} \quad R_3 = \frac{\langle K | V_k^C | D \rangle}{\langle K | V_k^L | D \rangle} \quad (14)$$

for which the KLM factors cancel out from numerator and denominator. Comparing the results of $R_2$ and $R_3$ (Fig. 5) to those of $R_1$ (Fig. 4), we see how the KLM prescription fails: The KLM corrected $R_1 \sim 0.75$ and $R_2 \sim 0.65$, although essentially independent of the quark mass, give different estimates for $Z_V$. The situation is even worse for the $Z_V$ estimate obtained from $R_3$. Thus, even at $\beta = 6.4$, there are big $\mathcal{O}(am)$ systematic errors unaccounted for by tree level KLM factors.

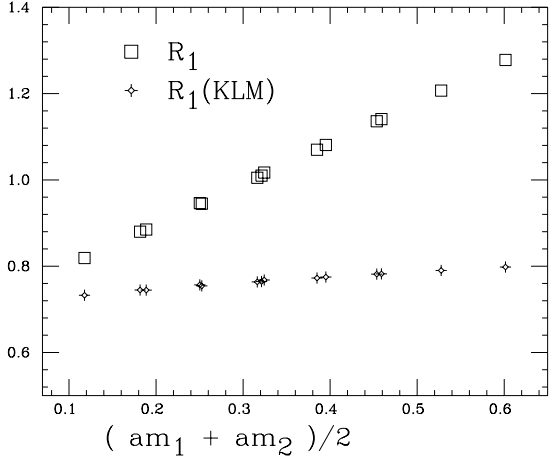

Figure 4. $Z_V$ calculated from $R_1$ without and with KLM corrections for Wilson fermions.

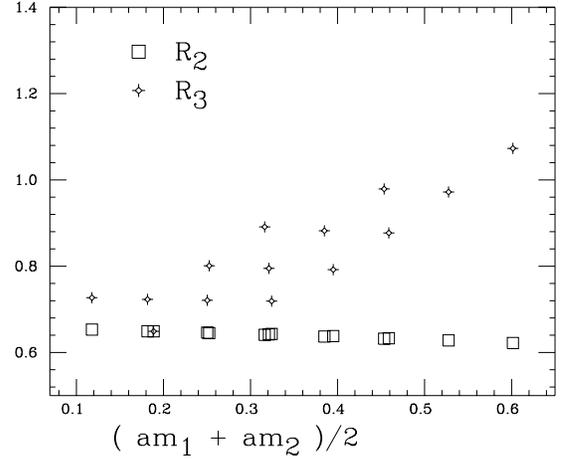

Figure 5. $Z_V$ calculated from $R_2$ and $R_3$ for Wilson fermions.

The same KLM ideas have been applied and tested in the Clover case. Here there are extra KLM factors arising from the Clover field rotations [10] $\psi \to \psi^R = \left[1 - \frac{a}{2}\gamma \cdot \vec{D}\right]\psi$. As an example we give the KLM factors of the quark correlation function of the local improved current $\langle 0|\psi_1^R(x) V_k^I(0) \bar{\psi}_2^R(y)|0\rangle$;

$$C_{CON}^{V_k^I} = \frac{2K_1(1+am_1)}{[1+\frac{1}{4}(1+am_1-\frac{1}{1+am_1})]^2}$$

$$\frac{2K_2(1+am_2)}{[1+\frac{1}{4}(1+am_2-\frac{1}{1+am_2})]^2} C_{LAT}^{V_k^I} \qquad (15)$$

Different KLM factors arise in the case of the conserved improved current $V_k^{CI}$ (see [22] for details).

Since the KLM factors respect Clover improvement, they are $1 + \mathcal{O}(a^2 m^2)$. Thus, they cannot rectify the $\mathcal{O}(g_0^2 am)$ corrections present in the Clover approach. This is demonstrated in Figs. 6, 7 ($\beta = 6.0$, $V = 16^3 \times 48$, 30 confs.) where we plot the ratios

$$R_4 = \frac{\langle 0|V_k^{CI}|\rho\rangle}{\langle 0|V_k^I|\rho\rangle} \quad and \quad R_5 = \frac{\langle K|V_k^{CI}|D\rangle}{\langle K|V_k^I|D\rangle} \qquad (16)$$

without and with KLM corrections [22]. The large linear $\mathcal{O}(g_0^2 am)$ term of the KLM corrected $R_4$ (Fig. 6) cannot be offset by the $1 + \mathcal{O}(a^2 m^2)$ KLM factor. In Fig. 7, the data does not fall on a smooth curve because $R_5$ also depends on $a^2(m_1 - m_2)^2$ (it is symmetric under $m_1 \leftrightarrow m_2$). Comparing the KLM corrected data of Figs 6 and 7, we see that the common linear behaviour up to $m \sim 0.5$ is followed by a non universal, ME-dependent behaviour at large quark mass. This systematic error cannot be corrected by KLM factors

## 8. $\Delta S = 2$

The reliability of the evaluation of the $B_K$ parameter with Wilson fermions has been plagued by the long standing problem of the bad chiral behaviour of the ME $\langle \bar{K}^0|O_{\Delta S=2}|K^0\rangle$. The non vanishing of the above ME in the chiral limit stems from two systematic effects. The first is the $\mathcal{O}(g_0^4)$ error in the 1-loop perturbative calculation of the logarithmically diverging RC, $Z_0$, and the finite constants, $Z_k$ ($k = 1, 2, 3$), which determine the mixing of the $O_{\Delta S=2}$ operator with dimension-6 operators of the wrong naive chirality:

$$O_{\Delta S=2}(\mu) = Z_0(a)[O_{\Delta S=2}(a) + Z_1(a)O_{SP}(a) \\ + Z_2(a)O_{VA}(a) + Z_3(a)O_{SPT}(a)] \quad (17)$$

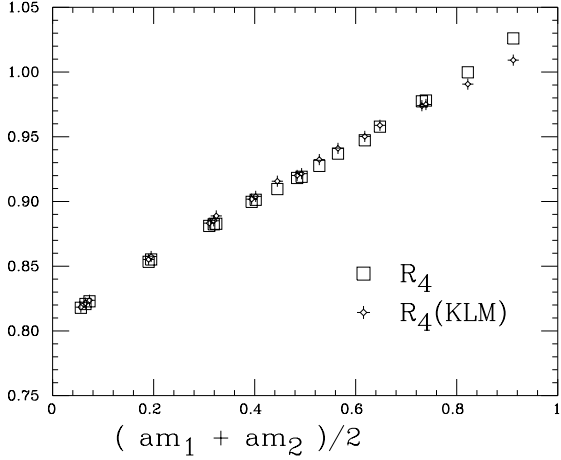

Figure 6. $Z_V$ calculated from $R_4$ without and with KLM corrections for Clover fermions.

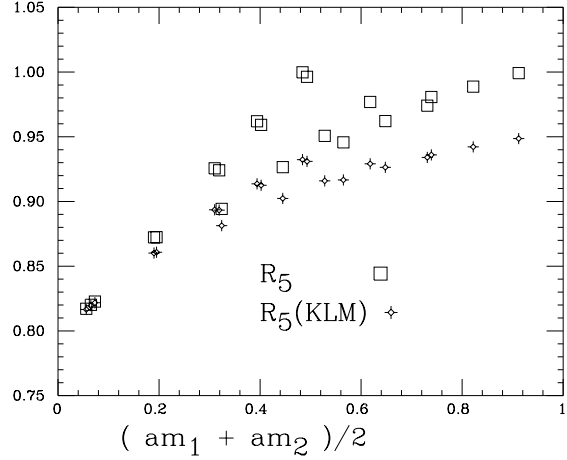

Figure 7. $Z_V$ calculated from $R_5$ without and with KLM corrections for Clover fermions.

(the definitions of the above operators can be found in [26]). The second systematic error comes from $\mathcal{O}(am)$ discretization effects in the four matrix elements $\langle \bar{K}^0|O_{\Delta S=2}|K^0\rangle$ and $\langle \bar{K}^0|O_k|K^0\rangle$ ($k = SP, VA, SPT$), which are calculated numerically at finite lattice spacing. In a preliminary comparative study of $B_K$ with Wilson and Clover fermions [24],[25], the latter source of error was investigated. The implementation of the Clover action caused a very marginal improvement of the chiral behaviour of the ME. Thus the principal source of error seems to be the perturbative determination of the RC's. In [26],[27] the NP method was applied to this operator (for details see [26]): The amputated 4-point quark correlation functions of the operators of eq.(17), suitably projected with projectors $\check{P}_{\Delta S=2}$ and $\check{P}_j$ ($j = 1, 2, 3$) are calculated numerically. By imposing the following four renormalization conditions on the subtracted correlation

$$\Lambda_s = \Lambda_{\Delta S=2} + Z_1\Lambda_{SP} + Z_2\Lambda_{VA} + Z_3\Lambda_{SPT} :$$

$$Tr[\check{P}_j\Lambda_s] = 0 \; ; j = 1, 2, 3$$

$$Z_{\Delta S=2}Z_\psi^{-2}Tr[\check{P}_{\Delta S=2}\Lambda_s]|_{p^2=\mu^2} = 1 \quad (18)$$

we obtain the four Z's. In Fig. 8 we report from [26],[27] a preliminary but very encouraging result (at $\beta = 6.0$ and $V = 16^3 \times 32$) with Clover fermions and NP mixing constants. It is clear that the NP mixing reproduces the expected chiral behaviour of the ME, whereas SPT and BPT do not. This is a crucial test that Wilson fermions have to pass if they are to predict the $B_K$ parameter with the same accuracy as staggered fermions.

## 9. Conclusions

For light quarks, a number of improvements (BPT, the Clover action, the NP method) have brought the different estimates of the RC's to close agreement. The newest encouraging results in this area come from the NP calculation of the RC's of the operators which determine the nucleon structure functions.

The NP estimates of the mixing coefficients of the $O_{\Delta S=2}$ operator apparently reproduce the expected chiral behaviour of the ME. If this preliminary result is verified by the more systematic runs currently performed by the APE group, it will reopen the way to the long awaited $\Delta I = 1/2$ calculation.

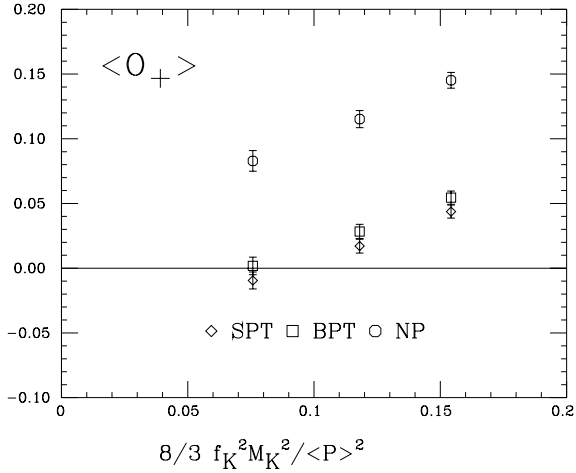

Figure 8. Chiral behaviour of $\langle O_+ \rangle = \langle \bar{K}^0|O_{\Delta S=2}|K^0\rangle/\langle P\rangle^2$ with mixing constants determined from SPT, BPT and NP methods. $f_K$ and $M_K$ are the decay constant and mass of $K^0$.

For charmed quark masses, WI estimates of RC's suffer from systematic discretization errors at the current $\beta$ values. These are only partially accounted for by tree level matching.

**Acknowledgements**

I am particularly indebted to M. Crisafulli, V. Lubicz, G. Martinelli, G. Rossi and C. Sachrajda, from and with whom I have learnt so much. Discussions with A. Kronfeld, S. Petrarca, C. Pittori, G. Schierholz and M. Testa are also gratefully acknowledged.